\documentclass[twocolumn,amsmath,amssymb, aps, pra]{revtex4-1}
\usepackage{graphicx}    
\usepackage{bm}
\usepackage{hyperref}    
\usepackage[mathlines]{lineno} 
\usepackage{xcolor} 
\usepackage[percent]{overpic}
\usepackage{setspace}
\usepackage{comment}
\usepackage{amsmath}

\begin{document}

\title{Wigner Function Reconstruction in Levitated Optomechanics}

\author{Muddassar Rashid}
\affiliation{Department of Physics and Astronomy, University
of Southampton, SO17 1BJ, United Kingdom}

\author{Marko Toro\v{s}}
\affiliation{Department of Physics and Astronomy, University
of Southampton, SO17 1BJ, United Kingdom}

\author{Hendrik Ulbricht} \email{hendrik.ulbricht@soton.ac.uk}
\affiliation{Department of Physics and Astronomy, University
of Southampton, SO17 1BJ, United Kingdom}

\date{\today}

\begin{abstract} 
We demonstrate the reconstruction of the Wigner function from marginal distributions of the motion of a single trapped particle using homodyne detection. We show that it is possible to generate quantum states of levitated optomechanical systems even under the effect of continuous measurement by the trapping laser light. We describe the opto-mechanical coupling for the case of the particle trapped by a free-space focused laser beam, explicitly for the case without an optical cavity. We use the scheme to reconstruct the Wigner function of experimental data in perfect agreement with the expected Gaussian distribution of a thermal state of motion. This opens a route for quantum state preparation in levitated optomechanics.

\end{abstract}


\maketitle

\section*{Introduction}
The realization of quantum features in the motion of massive objects is at the heart of many efforts in quantum science and technology. In order to demonstrate that a quantum state has indeed been prepared, sensitive measurement techniques have to be employed. Those are typically based on noise-cancelling homodyne techniques or sideband-resolved heterodyne methods \cite{Braginsky1995}. With such sensitive measurement tools at hand it is not unreasonable to even expect the observation of non-classical features in domains where one would typically not expect any quantum mechanics to be at work, such as at high temperatures \cite{purdy2017quantum} or for highly excited states \cite{Milburn2016}.

Quantum state reconstruction (QSR) is a common tool to analyse whether the state of a system is purely classical or comprises quantum features such as sub-Poissonian population statistics or quadrature squeezing \cite{Leonhardt2005}. The key ingredient for QSR is the ability to independently detect conjugate variables of the dynamics such as position, $x$ and momentum, $p$ of a continuous variable system such as a mechanical harmonic oscillator. Some QSR use the reconstruction of the so-called Wigner quasi-probability function in phase space, which has the distinct feature to show negative values if the observed state is non-classical, for instance a spatial superposition state. 

The Wigner function was first reconstructed experimentally for quantum states of light \cite{Smithey1993} and, thereafter, heavily used within quantum optics in the study and applications of the quantum nature of light. Today, the Wigner function has been also reconstructed for modes of molecular vibrations \cite{Dunn1995}, the motion of trapped atomic ions \cite{Wineland}, the spatial superposition state of atoms \cite{kurtsiefer1997measurement} and very recently for the thermal mechanical states of a harmonic oscillator in pulsed optomechanics \cite{Vanner2013}. QSR has also been applied to study both the effect of decoherence on a quantum system \cite{Deleglise2008} and the evolution of states relevant in quantum information processing \cite{Neergaard-Nielsen2006,Ourjoumtsev2007}.

The emerging research field of levitated optomechanics aims to study and control the motion of particles which are trapped in vacuum by light. Such levitated systems are well isolated from their environment, which dramatically reduces the effect of thermal noise on the centre-of-mass (\emph{cm})  motion of the trapped particle, as those can only weakly couple to its motion. In other words, extremely high quality factors of the mechanical oscillation of the particle in the trap can be achieved \cite{Chang2010,Gieseler2013}. As a consequence, levitated systems are
promising for manifold studies and applications such as macroscopic quantum superpositions \cite{Chang2010,Pflanzer,Bateman2014}, force sensing \cite{Ranjit2016,Hempston2017}, and single particle thermodynamics \cite{Gieseler2014,Millen,Erlangen2015}. Development in levitated optomechanics experiments over the last five years or so, has resulted in the successful demonstration of cooling  \cite{Kiesel2013,Millen2015,Jain2016,Vovrosh2017} to less than 100 phonons and squashing \cite{Rashid2016} the \emph{cm} motion of the particle.
\begin{figure}[t]
  \centering
  \includegraphics[width=1\linewidth]{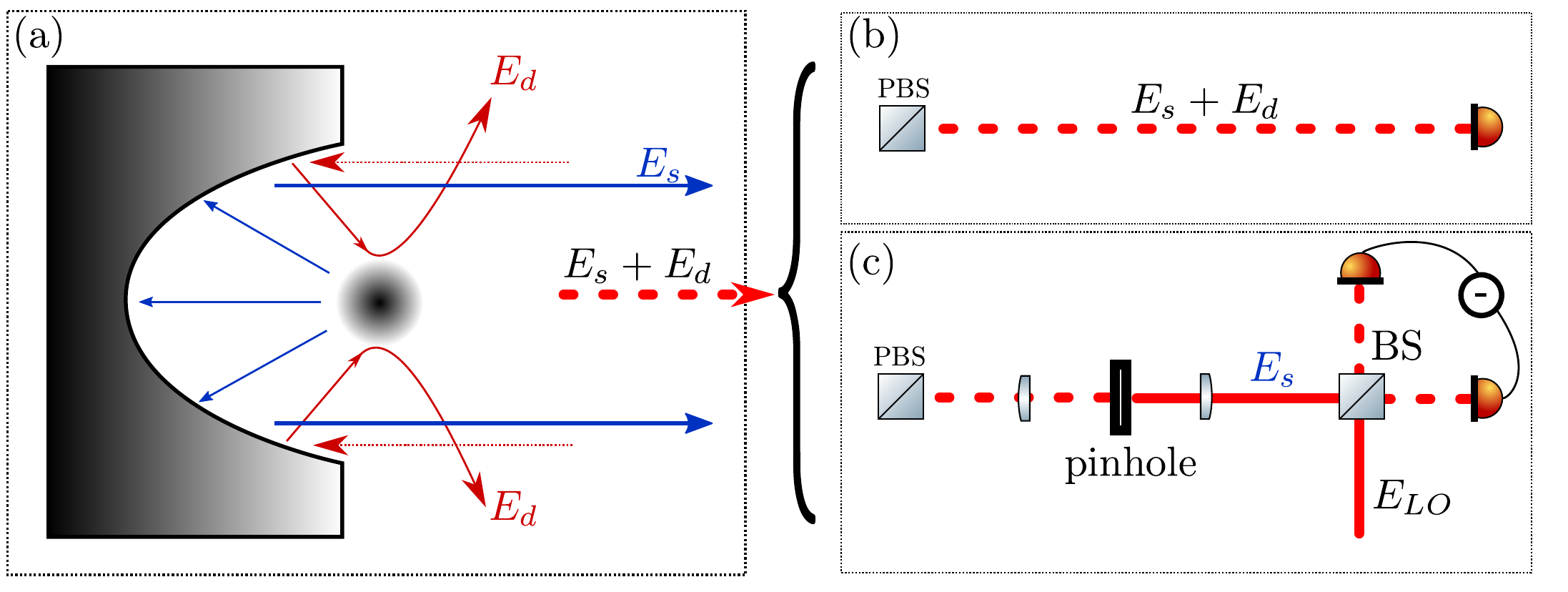}
  \caption{\textbf{Trapping \& Detection Schematics:} \textbf{(a)} The laser field is focussed by a parabolic mirror to a diffraction-limited spot, where a nanoparticle is trapped: we label the light field that generates the optical trap as $\bm{E}_{d}$. Part of $\bm{E}_{d}$ is Rayleigh scattered by the trapped particle, and approximately half is collected and collimated by the mirror: we label this field as $\bm{E}_{s}$. Using polarization optics the two fields, $\bm{E}_{s}$ and $\bm{E}_{d}$, are guided into the detection scheme: \textbf{(b)} {\bf the continuous homodyne (CH) detection}, where the interferometric fields are incident upon a single fast-photodetector, whilst in \textbf{(c)} {\bf the continuous balanced homodyne (CBH) detection}, $\bm{E}_{d}$ is spatially filtered using a pinhole, and $\bm{E}_{s}$ is directed through a 50:50 beamsplitter with a local oscillator, $\bm{E}_{LO}$, field into a balanced photodetector. $\bm{E}_{LO}$ is generated by splitting a small fraction of the trapping laser light before it is incident upon the parabolic mirror.}
\label{Fig:experimental_scheme}
\end{figure}

While there are manifold experimental and theoretical efforts on {\it how to generate} non-classical states of levitated optomechanics, here, we are concerned with {\it how to verify}  the successful generation of such states. To that end, we describe two homodyne detection methods for a levitated particle in a laser dipole trap that enable measurement of a particle's position and momentum independently and therefore to perform QSR. The methods are then used to reconstruct the Wigner function from experimental data in agreement with the expected result for a Gaussian thermal state. We leave the generation of truly quantum states to later research. 

\section*{Theoretical description}
The experimental setup we consider is shown in Fig. \ref{Fig:experimental_scheme}(a). A parabolic mirror is used to realise both the particle trap and the efficient collection of scattered light for the detection of the motion of the trapped particle. Importantly, the mirror enables an optical interferometric detection of the \emph{cm} motion of the trapped particle. That interference is between two light fields returning from the mirror: $\bm{E}_{s}$, which is Rayleigh scattered by the trapped particle and, $\bm{E}_{d}$, which is not scattered but diverges. 

In the following we describe theoretically the key elements for the desired QSR. In more detail, we quantize the electric field and the center of mass degree of freedom of the nanonoparticle. We then consider, the coupling between a polarizable point particle and the electric field. The light-matter coupling generates a trap, as well as an outgoing electric field due to Rayleigh scattering, which is used for quadrature measurement. We trace over the outgoing Rayleigh scattered light modes to obtain a Lindblad master equation to investigate the decoherence effect of continuous probing of the particle's position by the trapping laser. Specifically, we estimate the decoherence time for different superposition sizes, which we compare with the other relevant time-scales of the system. We consider two types of homodyne detection methods and under the assumption of the unperturbed harmonic evolution of the nanoparticle we construct the marginal distributions. We reconstruct the Wigner function using the inverse Radon transformation. In addition, we compare our free-space system with a cavity system.


{\bf Quantization of the electromagnetic field--}
In order to describe the coupling between the particle and the light, we consider a volume $V_q$ centered at the focal point of the parabolic mirror, where we quantize the electromagnetic field. Specifically, the focal point is in the center of the coordinate system, the $z$ axis is aligned with the optical axis and pointing away from the mirror, while the $x$ an $y$ axis are parallel to the transverse plane. 
We model the laser field $\hat{\bm{E}}_d$ as a Gaussian field:
\begin{equation}
\hat{\bm{E}}_d (\bm{r})=i E_0 \bm{\epsilon}_d \left( u (\bm{r}) \hat{a} - u (\bm{r})^* \hat{a}^\dagger \right),
\label{Ed}
\end{equation}
where
\begin{equation}
u(\bm{r})=\frac{w_0}{w(z)}e^{-\frac{x^2+y^2}{w(z)^2}} e^{i k z},
\label{udef}
\end{equation}
$E_0$ is the amplitude at the center of the beam waist, $\bm{\epsilon}_d$ is a transverse polarization vector, $w(z)=w_0\sqrt{1+(\frac{z}{z_R})^2} $, $w_0$ is the beam waist, $z_R=\frac{\pi w_0^2}{\lambda}$ is the Rayleigh range, $k=\frac{2\pi}{\lambda}$, $\lambda$ is the wavelength of the laser light and $\hat{a}$ ($\hat{a}^\dagger$) is the annihilation (creation) operator. The free electromagnetic field (the bath) is given by:
\begin{equation}
\hat{\bm{E}}_{f}(\bm{r})=i \sum_{\bm{k},\nu} \bm{\epsilon}_{\bm{k},\nu}  \sqrt{\frac{\hbar \omega_k}{2V_q \epsilon_0}} \left( v_{\bm{k}}(\bm{r}) \hat{a}_{\bm{k},\nu} - v_{\bm{k}}^*(\bm{r})\hat{a}^\dagger_{\bm{k},\nu} \right),
\label{Escatt} 
\end{equation}
where 
\begin{equation}
v_{\bm{k}}(\bm{r})=e^{i \bm{k} \cdot \bm{r}},
\label{vdef}
\end{equation}
$\epsilon_0$ is the permittivity of free space, $\bm{\epsilon}_{\bm{k},\nu}$ is the polarization vector, $\nu$ denotes the two independent polarizations, $\omega_k=c k$, $c$ is the speed of light and $\hat{a}_{\bm{k},\nu}$ ($\hat{a}_{\bm{k},\nu}^\dagger$) are the annihilation (creation) operators.

{\bf The optomechanical coupling--}
We also quantize the particle's center-of-mass degree of freedom: we denote the position operator as $\bm{\hat{r}}=(\hat{x},\hat{y},\hat{z})$ and the corresponding momentum operator as $\bm{\hat{p}}=(\hat{p}_x,\hat{p}_x,\hat{p}_z)$. We assume the following coupling between the dielectric particle and the electromagnetic field~\cite{Pflanzer,Pflanzer2012,Rodenburg2016}
\begin{equation}
\hat{H}_{\text{diel}}=-\frac{1}{2}\int_{V(\hat{\bm{r}})}d\tilde{\bm{r}} \hat{\bm{P}}(\tilde{\bm{r}}) \cdot \hat{\bm{E}}(\tilde{\bm{r}}),
\label{coupling1}
\end{equation}
where $V$ denotes the volume of the levitated particle,
\begin{equation}
\hat{\bm{E}}=\hat{\bm{E}}_d+\hat{\bm{E}}_f
\label{combined}
\end{equation}
is the total electric field and $\hat{\bm{P}}$ is the polarization vector. Specifically, we assume that~\cite{jackson1975electrodynamics}:
\begin{equation}
\hat{\bm{P}}=\epsilon_0 \epsilon_c \hat{\bm{E}},
\label{polarization}
\end{equation} 
where $\epsilon_c = 3 \frac{\epsilon_r-1}{\epsilon_r+2}$ and  $\epsilon_r$ denotes the dielectric function. 

Using Eqs.~\eqref{combined},~\eqref{polarization} in Eq.~\eqref{coupling1} we obtain three terms: (i) the trap potential ($\propto \hat{\bm{E}}^2_d$) ,(ii) the laser scattering term ($\propto \hat{\bm{E}}_f \cdot \hat{\bm{E}}_d$) and  (iii) the self-scattering of the free light field ($\propto \hat{\bm{E}}^2_f$). Among the two scattering terms we will only consider the dominant laser scattering term (ii), while we will neglect the self-scattering term (iii). For the present experiment, we can assume that field $\hat{\bm{E}}_d$ is strong and coherent, i.e. formally one can make the replacements $\hat{\bm{E}}_d\rightarrow \bm{E}_d $, $\hat{a}\rightarrow \alpha$ in Eq.~\eqref{Ed}, where $\bm{E}_d$, $\alpha$ are c-numbers. However, we will continue to use the operator notation, as the developed formalism could be also applied to other experiments, where the quantum nature of the field $\hat{\bm{E}}_d$ might be relevant. In addition, we assume that the fields $\hat{\bm{E}}_d$, $\hat{\bm{E}}_f$ are approximately constant over the volume of the particle and we make the formal replacement $\int_{V(\hat{\bm{r}})}d\tilde{\bm{r}} \rightarrow V \int \delta^{(3)} (\tilde{\bm{r}} - \hat{\bm{r}})d\tilde{\bm{r}}$ in Eq.~\eqref{coupling1}.  

\begin{figure*}[!ht]
  \centering
  \includegraphics[width = 0.85\linewidth]{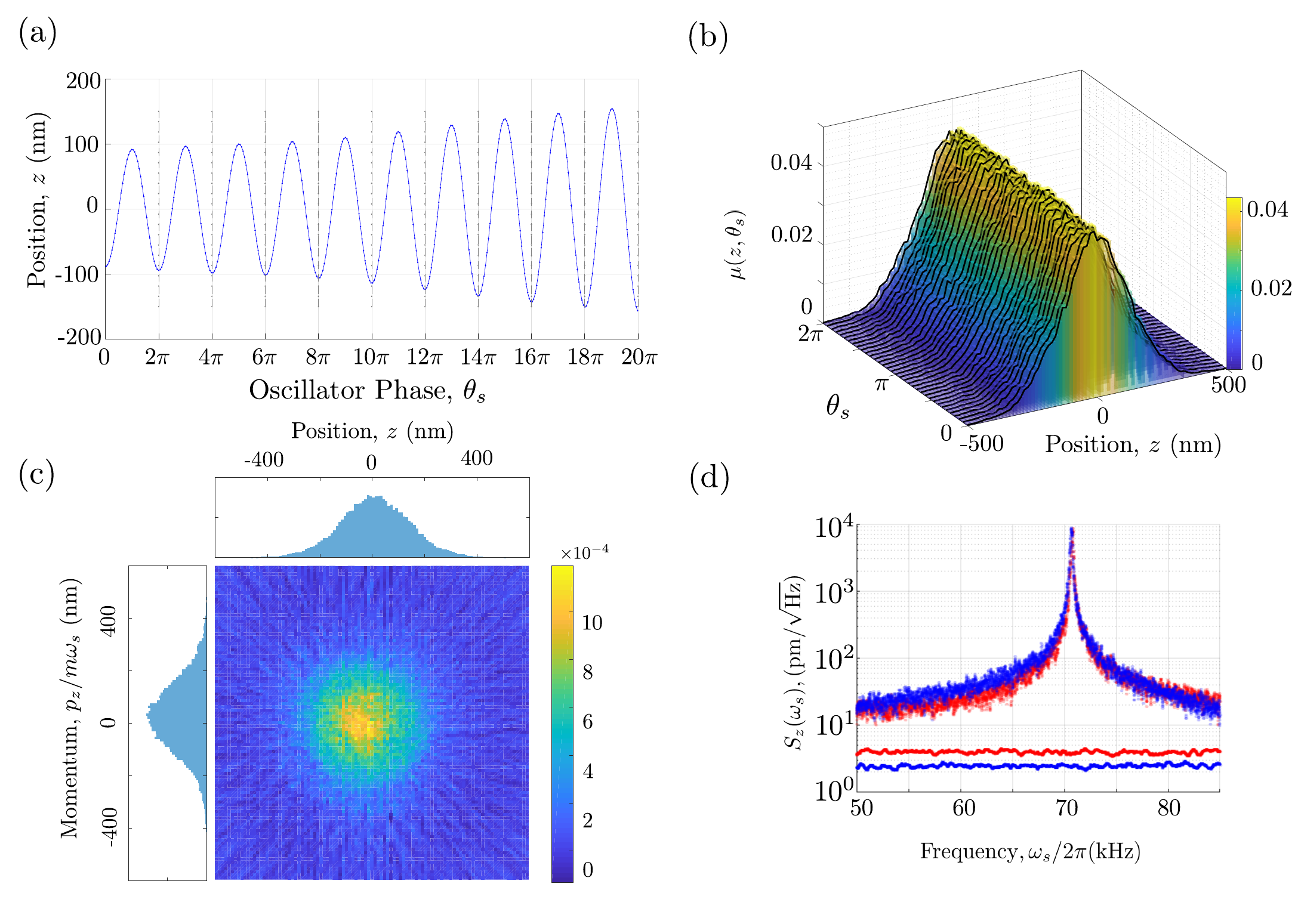}
  \caption{\textbf{Reconstructing the Wigner function from experimental data:} \textbf{(a)} Position signal of single trapped particle from the CBH detection. The vertical dashed-grey lines signify multiple $2\pi$ phases of the oscillator.  \textbf{(b)} The marginal distribution generated from (a) for the levitated
    particle and the associated \textbf{(c)} reconstructed Wigner
    function in phase space. \textbf{(d)} Power Spectral Density (PSD) from CH (red)
    and CBH (blue) methods.  Horizontal lines depict the respective
    noise floors in both detection schemes and show the slightly improved signal-to-noise ratio (SNR) for the CBH scheme. }
  \label{fig:ExpResults}
\end{figure*}

Let us now first discuss the trap potential term (i):
\begin{equation}
\hat{H}_{\text{trap}}=-\frac{\epsilon_0 \epsilon_c V}{2}  \hat{\bm{E}}_d(\hat{\bm{r}}) \cdot \hat{\bm{E}_d}(\hat{\bm{r}}),
\label{couplingT}
\end{equation}
Using Eq.~\eqref{Ed} in Eq.~\eqref{couplingT}, making the rotating wave approximation and considering only terms up to order $\mathcal{O}(\hat{\bm{r}}^2)$,
one obtains a harmonic trap~\cite{notec}:
\begin{equation}
U(\hat{\bm{r}})=2\epsilon_0\epsilon_c V E_0^2 |\alpha|^2 \left(\frac{\hat{x}^2}{w_0^2}+\frac{\hat{y}^2}{w_0^2}+\frac{\hat{z}^2}{2 z_R^2}\right).
\label{utrap}
\end{equation}
Combining Eq.~\eqref{utrap} with the nanoparticle's free evolution term one then obtains the Hamiltonian:
\begin{equation}
\hat{H}=\frac{\hat{\bm{p}}^2}{2m}+U(\hat{\bm{r}}),
\label{Hp}
\end{equation}
where $m$ is the mass of the nanoparticle.

{\bf Relevant aspects of decoherence theory for treatment of continuous measurement--}
Let us now discuss the laser scattering term (ii):
\begin{equation}
\hat{H}_{\text{Scatt.}}=-\epsilon_0 \epsilon_c V \hat{\bm{E}}_d(\hat{\bm{r}}) \cdot \hat{\bm{E}}_f(\hat{\bm{r}}),
\label{couplingS} 
\end{equation}
This term produces non-unitary dynamics for the center-of-mass. Specifically, assuming that the scattered photons have frequency $\omega_L$, applying the rotating wave approximation and tracing out the light degrees of freedom in the Born-Markov approximation, one obtains the following Lindblad term~\cite{nimmrichter2014macroscopic,stickler2016spatio}:
\begin{equation}
\begin{split}
\mathcal{L}[\hat{\rho}_t]=\gamma  \Big(&
\int d^2 n R(\bm{n})  u(\hat{\bm{r}}) v^*_{k\bm{n}}(\hat{\bm{r}}) \rho v_{k\bm{n}}(\hat{\bm{r}}) u^*(\hat{\bm{r}})\\
& -\frac{1}{2}\{|u(\hat{\bm{r}})|^2,\hat{\rho}\} \Big),
\end{split}
\label{lindblad}
\end{equation}
where $\hat{\rho}_t$ is the center-of-mass statistical operator, $R(\bm{n})=\frac{3 \sin^2 (\theta)}{8\pi}$ is the angular distribution of a radiating dipole, $\theta$ is the polar angle, $\bm{n}$ is a unit vector and $u$, $v$ are defined in Eqs.~\eqref{udef},~\eqref{vdef}, respectively. The scattering rate is given by:
\begin{equation}
\gamma=\frac{\sigma}{\pi w_0^2}\frac{P}{\hbar \omega_L},
\label{eq:gamma}
\end{equation}
where $\sigma=\frac{8\pi^3 \epsilon_c V^2}{\lambda^4} $ is the Rayleigh cross section, $P$ is the laser power and $\omega_L=\frac{2\pi c}{\lambda}$ is the laser light frequency.

Thus, combining the contribution from the terms (i) and (ii), one obtains the following dynamics for the center-of-mass of the nanoparticle:
\begin{equation}
\frac{d\hat{\rho_t}}{dt}=-\frac{i}{\hbar}[\hat{H},\hat{\rho}_t]+\mathcal{L}[\hat{\rho}_t],
\label{master_eq}
\end{equation}
where $\hat{H}$ and $\mathcal{L}$ are defined in Eqs.~\eqref{Hp} and \eqref{lindblad}, respectively. The classical gradient force and radiation pressure force~\cite{harada1996radiation} can be re-obtained in this formalism as $\bm{F}_\text{Grad.}= -\langle \frac{\partial \hat{U}}{\partial\hat{\bm{r}}}\rangle$ and  $\bm{F}_\text{Scatt.}= \langle\hat{\bm{p}}\mathcal{L}[\hat{\rho}] \rangle$, respectively~\cite{nimmrichter2014macroscopic}. The radiation pressure force $\bm{F}_\text{Scatt.}$ creates a small offset of the equilibrium position of the harmonic potential for the mechanical degree of freedom, which we will neglect in the discussion~\cite{noteb}.

{\bf Output fields, which are used for measurement of the particle's position--}
We have thus far discussed the dynamics of the nanoparticle. We will now consider the scattered light field, which after being reflected by the mirror, travels towards the detector: this field, as anticipated above, will be used to reconstruct the state of the nanoparticle.

To ease the following analysis, we consider only the motion along the $z$-axis. In this case, 
the Hamiltonian in Eq.~\eqref{Hp} reduces to $\hat{H}=\hbar \omega_s \hat{b}^\dagger\hat{b}$, where $\hat{b}=\sqrt{\frac{1}{2m\hbar\omega_s}} (m\omega_s \hat{z} + i \hat{p}_z)$,  
$\omega_s=\sqrt{\frac{2 \epsilon_c P }{\rho c \lambda z_R^3 }}$ and $\rho$ is the particle density (we have used $\epsilon_0 E_0^2|\alpha|^2 =\frac{P}{ c \pi w_0^2}$). Moreover, we assume that the scattered electric field has the same polarization and opposite wave-vector as the incoming electric field. In this case, setting $\hat{x}=0$ and $\hat{y}=0$, the mode functions in Eqs.~\eqref{udef} and~\eqref{vdef} simplify to $u=e^{ikz}$ and $v=e^{-ikz}$, respectively. To simplify the notation we denote the annihilation operator associated to the latter mode function as $\hat{c}$.
Using then Eqs.~\eqref{Ed}, \eqref{Escatt}, \eqref{couplingS}, neglecting the fast rotating terms, we obtain~\cite{notec}: 
$\hat{H}_{\text{Scatt.}}=-C (e^{2ik\hat{z}} \alpha \hat{c}^\dagger+\text{H.c}) $,
where $C=\epsilon_0 \epsilon_c V E_0 \sqrt{\frac{\hbar \omega_L}{2V_q \epsilon_0}}$. We now suppose that the scattered mode is initially unpopulated (before the incoming field is scattered) and that the evolution is given only by the term $\hat{H}_{\text{Scatt.}}$. Moreover, we assume the mechanical degree of freedom remains unchanged during the time of the interaction. It is then straightforward to obtain $\hat{c}_{t+\tau}=\frac{iC \alpha\tau}{\hbar} e^{2ik\hat{z}_t}$, where $t$ is the initial time, $\tau$ is the interaction time and we have explicitly written the time-dependence.  

We now consider the electric fields that are scattered in other directions (but are still reflected by the mirror and collimated towards the detector). Specifically, we suppose that during the interaction with the nanoparticle such a field acquires the same phase factor $e^{2ik\hat{z}}$. We note that this field, once reflected by the mirror, has approximately the same polarization as the field $\hat{\bm{E}}_d$. We denote this reflected field as $\hat{\bm{E}}_s$ (see Fig.~\ref{Fig:experimental_scheme}(a)). To describe the fields near we detector we project on the polarization vector $\bm{\epsilon_}d$, i.e. $\hat{E}_{d}=\bm{\epsilon_d}\cdot \hat{\bm{E}}_{d}$ and $\hat{E}_{s}=\bm{\epsilon_d}\cdot \hat{\bm{E}}_{s}$. We decompose them in full generality as~\cite{notec}:
\begin{align}
\hat{E}_{d}=&\text{Re}\left( A e^{i(\phi_d-\omega_L t)}\right)\\
\hat{E}_{s}(\hat{z}_t)=&\text{Re}\left(  B e^{i(\phi_s+2k\hat{z}_t-\omega_L t)}\right), \label{Es}
\end{align}
where $\text{Re}$ denotes the real part and $A$, $B$, $\phi_d$, $\phi_s$   are real numbers. 

{\bf Description of detection schemes--}
We have considered two different detection schemes, which are graphically depicted in Fig.~\ref{Fig:experimental_scheme} (b) and (c): we will refer to them as continuous homodyne (CH) and continuous balanced homodyne (CBH) detections, respectively. In a nutshell, these two detection schemes rely on the system's unperturbed harmonic evolution to generate rotations of the quadratures in phase space~\cite{leonhardt1996observation}.
 
Consider, first, the CH detection scheme. The signal intensity is given by:
\begin{equation}
   \hat{I}(\hat{z}_t) =  c\epsilon_0\mathbb{E}\left[(\hat{E}_{d}+\hat{E}_{s}(\hat{z}_t)) ^2\right],  \label{sl}
\end{equation}
where $\mathbb{E}$ denotes the temporal average over the interval $\frac{2\pi}{\omega_L}$.
The number of photons hitting the detector in the time interval $[t_i,t_i+T )$ is given by~\cite{gardiner2004quantum}:
\begin{equation}
  \hat{N}(\hat{z}_{t_i})=\frac{\sigma_d}{\hbar \omega_L} \int_{t_i}^{t_i+T}  \hat{I}(\hat{z}_t)   dt,
  \label{N}
\end{equation}
where $T$ is the (constant) integration time and $\sigma_d$ is the area of the detector. As the notation already suggests, we assume that the the mechanical evolution is much slower than the duration of a single recording. Moreover, we assume that the evolution of the mechanical degree of freedom is given by $\dot{\hat{b}}=-i\omega_S\hat{b}$, on the time-scale of data recording. Thus, using
Eqs.~\eqref{sl}, \eqref{N} we obtain:
\begin{equation}
  \hat{N}(\hat{z}_{t_i}) = \frac{c\epsilon_0\sigma_d T}{2\hbar \omega_L}( A^2+B^2+2 A B \cos (\Delta\phi-2k\hat{z}_{t_i})), \label{Ns}
\end{equation}
where $\Delta\phi=\phi_d-\phi_s$. We now assume $|\Delta\phi -\pi n| \gg |2k\hat{z}_{t_i}|$, with $n\in \mathbb{Z}$, and Taylor expand to obtain a simplified expression:
\begin{equation}
  \hat{N}(\hat{z}_{t_i}) = C_1 + C_2  + D \hat{z}_{t_i}, \label{Ns_approx}
\end{equation}
where $C_1=\frac{c\epsilon_0\sigma_d T}{2\hbar \omega_L}(A^2+B^2)$, $C_2=\frac{c\epsilon_0\sigma_d T}{\hbar \omega_L}  A B \cos (\Delta\phi)$ and $D=\frac{c\epsilon_0\sigma_d T}{\hbar \omega_L}k AB \sin (\Delta\phi) $.

Let us now consider the CBH detection scheme. We now have two signal intensities:
\begin{align}
 \hat{I}_1(\hat{z}_{t}) &=  c\epsilon_0\mathbb{E}\left[(\hat{E}_{d}+\hat{E}_{s}(\hat{z}_{t}))^2\right],\label{sl2n} \\
 \hat{I}_2(\hat{z}_{t}) &=  c\epsilon_0\mathbb{E}\left[(\hat{E}_{d}-\hat{E}_{s}(\hat{z}_{t}))^2\right].
 \label{sl2}
\end{align}
The two signals are integrated by the respective
detectors $1$, $2$ in the time interval $[t_i,t_i+T )$ and then
subtracted~\cite{gardiner2004quantum}:
\begin{equation}
  \hat{N}(\hat{z}_{t_i})=\frac{\sigma_d}{\hbar \omega_L} \int_{t_i}^{t_i+T} \left(\hat{I}_1(\hat{z}_t)- \hat{I}_2(\hat{z}_t)\right) dt,
  \label{Ndiff}
\end{equation}
where $T$ is the (constant) integration time and $\sigma_d$ is the area of each detector. Starting from Eq.~\eqref{Ndiff}, after a similar calculation as the
one for the CH detection scheme, we obtain in place of Eq.~\eqref{Ns}
\begin{equation}
  \hat{N}(\hat{z}_{t_i}) = \frac{c\epsilon_0\sigma_d T}{\hbar \omega_L}  2 A B  \cos (\Delta\phi-2k\hat{z}_{t_i}) 
  \label{Ns2}
\end{equation}
and in place of Eq.~\eqref{Ns_approx}:
\begin{equation}
  \hat{N}(\hat{z}_{t_i}) = 2 C_2  + 2 D \hat{z}_{t_i}, \label{Ns_approx3}
\end{equation}

We can exploit Eqs.~\eqref{Ns_approx} or \eqref{Ns_approx3} to devise state
reconstruction methods. Specifically, the method, which we have
investigated experimentally in this paper, is to consider a single
system and continuously measure the quadrature $\hat{z}_{t_i}$ at
different times $t_i$. The free evolution of a harmonic oscillator for
a time $t_i$ corresponds to a rotation in phase space of angle
$\omega_s t_i$. Thus, the measurements of $\hat{z}_{t_i}$ can be used
to construct the marginals $\mu(z;\theta_s)$~\cite{Leonhardt2005},
where $\theta_s=\omega_s t_i\mod 2\pi$. This method, where we consider
a single system and continuous measurements of the quadrature, could
be combined with the quantum-state sampling
method~\cite{d1994detection,d1994precision,kuhn1994determination}: the
reconstructed state is updated, increasing the accuracy of the
reconstruction, by each new recorded value of the quadrature. However, in this paper, we adopt the inverse Radon transformation to reconstruct the Wigner function \cite{lee2012phase}.

\section*{Experimental implementations and results}
We now apply the theory developed in the previous section to reconstruct the Wigner function of a thermal state of motion and discuss the decoherence effect of continuous measurement from the trapping laser. 

The motion of the particle in the trap is measured by the detection of the intensity of light according to Eqs.~(\ref{sl}) and (\ref{sl2n}), (\ref{sl2}) for CH and CBH, respectively. We realise the setup in Fig. \ref{Fig:experimental_scheme}(a), by using 1550 nm laser light of $650$ mW, incident upon the parabolic mirror made of aluminium, with an aperture of $3$ mm. At the diffraction limited focal spot generated by the mirror, we trap a silica particle of diameter $34$ nm, which oscillates with $\omega_s = 2\pi \times 70$ kHz, where $\omega_s$ is the frequency of the $z$-motion. To realise both the detection schemes in Fig. \ref{Fig:experimental_scheme}(b)(c), we use the same type of photodetectors with a bandwidth of 4 MHz. For both CH and CBH schemes we record the intensity at the detectors for one second (see Eqs.~\eqref{Ns_approx} and \eqref{Ns_approx3}, respectively). In post-processing, this signal is filtered for the $z$ degree of freedom of the particle and converted from time $t$ to oscillator phase, $\theta_s = \omega t$ mod $2\pi$, where $\omega_s$ is the oscillator's frequency, as shown in Fig.~\ref{fig:ExpResults}(a). 

This measured signal contains the information of the evolution of the particle's harmonic motion, and we in effect sample each oscillation phase thousands of times during a one second time trace measurement. We collect the intensity value for each phase of the oscillating particle for many oscillation cycles, which gives the statistics of the measurement. The resulting marginal distributions $\mu(z,\theta_s)$ are generated for phases 0 to $2\pi$, as shown in Fig. \ref{fig:ExpResults}(b). We use the convention that when $\theta_s = 0$ we acquire the marginal distribution of position $\mu(z,0) = \mu(z)$, while for $\mu(z,\pi/2) = \mu(p_z)$, we get the marginal for the momentum $p_z$. This is an important point as it allows for the extraction of both quadratures independently in order to reconstruct the state and is at the heart of each homodyne detection. By using all marginal distributions and applying the inverse Radon transformation \cite{lee2012phase} we obtain the Wigner function of the thermal state of motion of the particle, as shown in Fig. \ref{fig:ExpResults}(c). Both Fig. \ref{fig:ExpResults}(b)(c) show the state distribution is Gaussian and centred about the origin of phase space. This distribution is the expected result for a trapped particle in thermal equilibrium with the environment. Especially, we do not expect any negative value of the Wigner function for a thermal state. 

As a further result, Fig. \ref{fig:ExpResults}(d), shows the power spectral density (PSD) obtained from both the CH and the CBH detection schemes at $1 \times 10^{-2}$ mbar and $T=300K$ for a particle of radius $34$ nm. The size of the particle has been extracted from a Lorenzian fit to the PSD. Although, both signal peaks are the same, the noise floor has decreased by a factor of
two for the CBH scheme if compared to CH, implying a better signal-to-noise ratio (SNR) for CBH, which is promising for future work on state preparation and cooling. However, the CBH scheme, is very sensitive to any phase changes in the LO arm and for the experiments reported here, a laser pointing instability over time scales of a few seconds causing variation in the detected light intensity. This is the identified technical limit on the SNR in the present experiments, as it limits the detection integration time. This SNR has to be improved down the shot noise limit in future experiments to detect non-classical states, which has already been demonstrated in optomechanical systems \cite{clerk2010quantum,purdy2013observation,aspelmeyer2014cavity}.

\begin{figure}[t]
  \centering
  \includegraphics[width=1\linewidth]{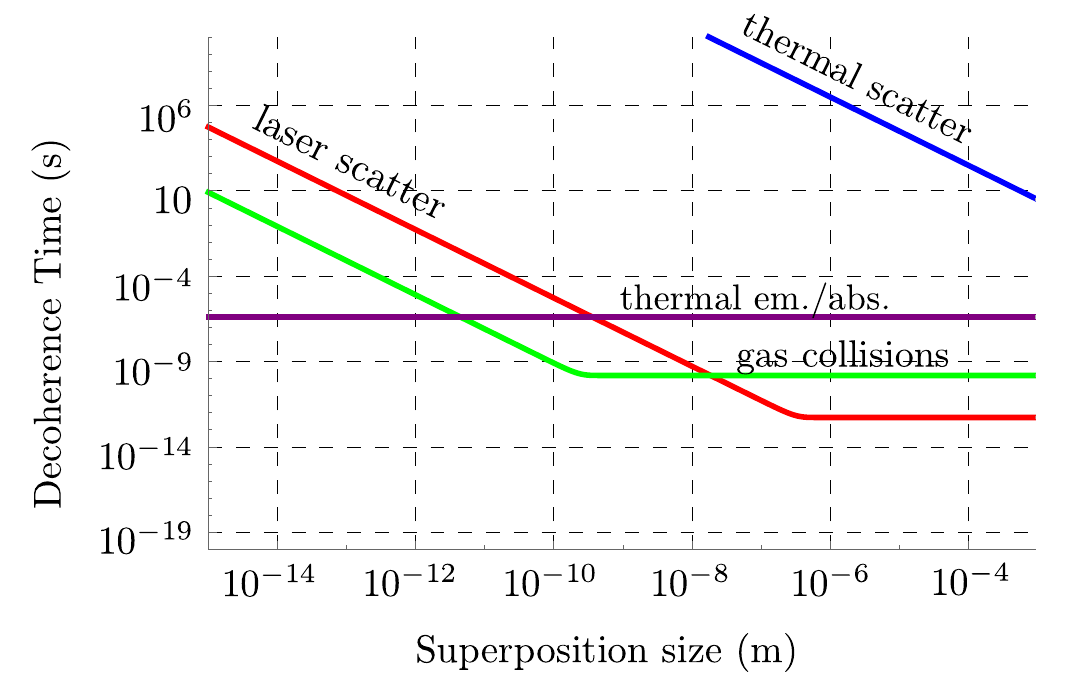}
  \caption{\textbf{Decoherence times from continuous measurement by the trapping laser for the example of a superposition state:} The decoherence time due to laser light scattering (red line) is estimated using Eq.~\eqref{join}. We have plotted for comparison the decoherence time due to emission and absorption of thermal photons (purple line), where we have assumed that the internal and external temperatures are the same, namely at 300 K, scattering of thermal photons (blue line)  and gas collisions (green line)  at a background gas pressure of $10^{-2}$mbar \cite{hornberger2004theory,romero2011quantum,carlesso2016decoherence}.}
    \label{Fig:decoherence2}
\end{figure}

Invariably, the {\it continuous probing} from the laser field yields a strong limit on the coherence times of non-classical correlations. We treat the continuous probing as a decoherence effect and derive a master equation from Eq.~(\ref{master_eq}), in the limit of small displacements. Setting $\hat{x}=0$ and $\hat{y}=0$ and expanding to quadratic order in $\hat{z}$, one obtains a simplified master equation:
\begin{equation}
  \frac{d\hat{\rho}}{d t}=-\frac{i}{\hbar}[\hat{H}_p,\hat{\rho}_t]-\Gamma  [\hat{z}, [\hat{z},\hat{\rho}]],
\end{equation}
where $\Gamma=\frac{12 \pi^2\gamma}{5\lambda^2}$. The quantities
$\gamma$, defined in Eq.~\eqref{eq:gamma}, and $\Gamma$ govern the short and long wavelength limits,
respectively. We can define an effective decoherence time $\tau$, function of
the superposition size $\Delta z$, which joins the two regimes (see
Fig.~\ref{Fig:decoherence2})~\cite{hornberger2004theory,romero2011quantum,
carlesso2016decoherence}:
\begin{equation}
\tau(\Delta z)=\left(\gamma \tanh (\Gamma\Delta z^2/\gamma )\right)^{-1}.
\label{join}
\end{equation}
For a spatial superposition of size $\Delta z \sim 0.1 \text{nm}$ the resulting decoherence time is estimated to be on the order of 10 $\mu$s, as plotted in Fig.~\ref{Fig:decoherence2}. This time corresponds to one full oscillation period of the present levitated system. More generally, assuming we would suppress all other sources of decoherence and prepare a spatial superposition state of size $\Delta z$, that state would persist for a time $\tau(\Delta z)$.

This result sets the scale for any possible generation of non-classical correlations, specifically, to observe negativity in the Wigner reconstruction. However, as can be seen from Fig.~\ref{Fig:decoherence2}, for the current experiment, the decoherence due to other environmental sources becomes relevant at these distances as well. Not surprisingly, decoherence due to gas collisions with background gas at the given pressure of $10^{-2}$mbar is the dominating effect by some orders of magnitude, but also the effect of emission and absorption of thermal (black body) photons by the trapped particle at 300 K shows a stronger decoherence effect compared to the continuous trapping laser light scattering by the trapped particle.

\section*{Discussion and Conclusion}
The state reconstruction presented in this paper relies on the assumption of the unperturbed harmonic evolution of the trapped system. Any future manipulation scheme to generate non-classical states in conjunction with our reconstruction method, has to take this harmonic motion assumption into account and therefore has to be on shorter timescales compared to the oscillation. This assumption is satisfied on the time-scales of data recording if (i) the coupling to the environment (background gas particles and thermal photons) is small and (ii) the decoherence due to scattering of photons from the trapping laser is sufficiently weak. While for a classical system these two assumptions can be relaxed, they become of crucial importance in order to detect non-classicality. On the one hand, decreasing the pressure and temperature of the environment, as well as lowering the internal temperature of the trapped particle~\cite{rahman2017laser}, will isolate the system well enough to satisfy condition (i). However, the action of continuous quadrature measurements, which are not of the quantum non-demolition (QND) type, is ultimately constrained by Heisenberg's uncertainty
principle~\cite{braginsky1980quantum}.

A possible solution to satisfy both (i) and (ii) is to consider an ensemble of identically prepared systems~\cite{Pflanzer}. This could be achieved in the following way: We apply the construction protocol
$C$ that constructs the desired state from an arbitrary initial state. We then take quadrature measurements, which we call protocol $M$, for only a time $\beta \omega_S^{-1}$, where $\beta$ is a positive number.  We can now repeat the protocols $C$ and $M$, gradually constructing the marginal distributions $\mu_{\theta}(z;\theta)$, where $\theta=\omega_s t^{(j)}$ and $0\leq t^{(j)}<\beta \omega_S^{-1}$ denotes the time from the start of the $j$-th time we apply the protocol $M$. In this way we have to satisfy (i) and (ii) only for the time $\beta \omega_S^{-1}$, which makes the method more experimentally feasible. We leave a more rigorous
analysis of the limits of validity of the proposed detection schemes, for future research.

The CH and CBH detection methods are to be compared with the detection scheme, which is usually adopted in cavity optomechanics. There, one creates an ensemble of identically prepared systems for each phase $\phi$ of the LO and measures the marginal distributions $\mu_{\phi}(\delta z;\phi)$ corresponding to the rotated quadrature $\exp(i\delta\hat{b}^\dagger\delta\hat{b}\phi)\delta\hat{z}\exp(-i\delta\hat{b}^\dagger\delta\hat{b}\phi)$, where $\delta\hat{b}$ and $\delta\hat{z}$ denote the fluctuations around the the steady state values~\cite{vitali2007optomechanical}. From the marginal distributions one can then, at the end of data collection, reconstruct the state of the system~\cite{raymer1995ultrafast}. This method has the drawback that the algorithm for state reconstruction is applied at the very end of data collection, a feature which makes it unappealing for state control and manipulation.  Moreover, one has to experimentally control the phase of the LO, which is non-trivial.

However, there is more natural relation between cavity and free space systems. Consider first the free space system: the strong coherent light field scatters off the trapped nanoparticle and then, without any further interaction, is reflected by the mirror towards the detector. A lossy cavity system, where the cavity mode is initially unpopulated and driven by an a strong coherent light field, behaves in a similar way. Moreover, one can show that the CH and CBH detection schemes can be used also for cavity systems. In addition, one obtains the following correspondence $\frac{g_0}{\kappa}= 2\pi\frac{z_{\text{zpf}}}{\lambda}$, where, on the left hand-side, $g_0$ and $\kappa$ are the opto-mechanical coupling and the optical decay rate for a cavity system, and, on the right hand-side, $z_{\text{zpf}}$ and $\lambda$ are the quantities for free spaces system discussed here above (see Appendix \ref{sec:Appendix}).

In conclusion, we have shown that the parabolic mirror setup used in levitated optomechanics can be utilised for carrying out homodyne detection. We present a theory on how momentum and position of the mechanical oscillator are extracted from photon counts at the detector. We have also shown that by utilising the evolution of the particle in the harmonic trap we can track its phase, which can be
used to generate the marginal distributions. By applying an inverse Radon transform to the marginals, we carry out Wigner reconstruction for a thermal state of a levitated nanoparticle.  Thus, we have
demonstrated detection techniques fit to prove non-classical features of levitated optomechanics in parabolic mirror traps, future work will aim to complement this by the generation of such non-classical motional states. For example, the state detection and manipulation could be implemented using the Kalman filter~\cite{kalman1960new,kalman1961new}.

{\bf Acknowledgements--}
We would like to thank Markus Rademacher, Ashley
Setter, Chris Timberlake and George Winstone for discussions. We wish
to thank for funding, The Leverhulme Trust and the
Foundational Questions Institute (FQXi).

\bibliographystyle{apsrev4-1}
\bibliography{wigner}

\begin{thebibliography}{58}
\bibitem [{not({\natexlab{a}})}]{notec}%
  \BibitemOpen
  \href@noop \bibinfo {We make the replacement
  $\hat{\bm{E}}_d\rightarrow \bm{E}_d $ and $\hat{a}\rightarrow \alpha$ in
  Eq.~\eqref{Ed}, where $\bm{E}_d$, $\alpha$ are c-numbers.}\BibitemShut
  {Stop}%
\bibitem [{not({\natexlab{b}})}]{noteb}%
  \BibitemOpen
  \href@noop  \bibinfo {The laser light
  travelling towards the mirror creates an analogous radiation pressure force,
  which we will also neglect.}\BibitemShut {Stop}%
\bibitem [{not({\natexlab{c}})}]{notex}%
  \BibitemOpen
  \href@noop \bibinfo {We first apply this
  transformation to the Schr{\"o}digner picture and then switch to the
  Heisenberg picture.}\BibitemShut {Stop}%
\end{thebibliography}

\clearpage

\appendix 
\renewcommand{\theequation}{A.\arabic{equation}}
\setcounter{equation}{0}
\section{Detection schemes in Cavity optomechanics}
\label{sec:Appendix}
It is instructive to make a comparision between a free-space and a cavity opto-mechanical setup~\cite{leuchs2013light}. In this appendix, we show that the free-space system, considered in the main text, is analogous to a lossy cavity opto-mechanical system. Specifically, we consider the following Hamiltonian for a cavity opto-mechanical system~\cite{law1995interaction,paternostro2006reconstructing,vitali2007optomechanical,aspelmeyer2014cavity}:
\begin{equation}
\begin{split}  \hat{H}=&\hbar\omega_S\hat{b}^\dagger\hat{b}+ \hbar\omega_L \hat{a}^\dagger\hat{a}- \hbar \frac{g_0}{z_{\text{zpf}}} \hat{a}^\dagger\hat{a} \hat{z} \\
&+i\hbar E (\hat{a}e^{i\omega_L t}-\hat{a}^\dagger e^{-i\omega_L t}),
\end{split}
  \label{Hamiltonian}
\end{equation}
where $\hat{b}$ ($\hat{b}^\dagger$) denotes the mechanical
annihilation (creation) operator,  $\omega_s$ is the mechanical frequency, $\hat{a}$
($\hat{a}^\dagger$) denotes the cavity annihilation (creation) operator, $g_0=-z_{zpf}\frac{\partial \omega_{\text{cav}}}{\partial z}|_{z=0}$ is the opto-mechanical coupling, $\omega_{\text{cav}}(z)$ is the cavity frequency, $\hat{z}=z_{zpf}(\hat{b}+\hat{b}^\dagger)$, $z_{zpf}=\sqrt{\frac{\hbar}{2m\omega_s}}$, $\omega_s$ is the frequency of the mechanical degree of freedom, $m$ is the mass of the mechanical degree of freedom, $E=\sqrt{\frac{2 P\kappa}{\hbar \omega_L}}$, $P$ is the laser power, $\omega_L=\omega_{\text{cav}}(0)$ and $\kappa$ is the cavity decay rate. Note that this cavity system has zero laser detuning, i.e. $\Delta =\omega_L-\omega_{\text{cav}} (0)=0$.

We write the corresponding (non-linear) Langevin equations in the interaction picture with respect to $\hbar \omega_L\hat{a}^\dagger\hat{a}$~\cite{notex}:
\begin{align}
  \dot{\hat{a}}=&-\frac{\kappa}{2}\hat{a}
                         +i\frac{g_0}{z_{\text{zpf}}} \hat{z}\hat{a}  +E+\sqrt{\kappa}\hat{a}_{\text{in}}, 
                         \label{adot}\\
  \dot{\hat{b}}=&-\left(i\omega_S+\frac{\xi}{2}\right)\hat{b}
                  +ig_0\hat{a}^\dagger \hat{a} + \sqrt{\gamma}\hat{b}_{\text{in}},	
   \label{bdot}
\end{align}
where $\xi$ is the mechanical decay rates and  $\hat{a}_{\text{in}}$, $\hat{b}_{\text{in}}$ are input noise operators~\cite{gardiner1985input}.

{\bf Relation between free space and cavity--}
We now write explicitly the time-dependence. The solution to Eq.~\eqref{adot} is given by:
\begin{equation}
\hat{a}_t=e^{-\frac{\kappa}{2}t}\hat{U}_t \hat{a}_0 + E e^{-\frac{\kappa}{2}t} \hat{U}_t
\int_0^t e^{\frac{\kappa}{2}\tilde{s}} \hat{U}_{\tilde{s}}^\dagger d\tilde{s},
\label{alpha1}
\end{equation}
where 
\begin{equation}
\hat{U}_t=\mathcal{T} \left( e^{i \frac{g_0}{z_{\text{zpf}}}\int_0^t \hat{z}_s ds} \right),
\label{Udefinition}
\end{equation}
$\mathcal{T}$ denotes the time-ordering operator and $\hat{a}_0$ is the initial value. The factor $e^{-\frac{\kappa}{2}(t-\tilde{s})}$ in Eq.~\eqref{alpha1} constrains the $\tilde{s}$ integration to the interval $[t-\frac{2}{\kappa},t]$, where we assume that $\hat{z}$ does not evolve significantly. We make the approximation:
\begin{equation}
\int_0^{\tilde{s}} \hat{z}_s ds=\int_0^{t} \hat{z}_s ds-\hat{z}_t (t-\tilde{s}). 
\end{equation}
Specifically, using this relation and Eq.~\eqref{Udefinition}, we have:
\begin{equation}
\hat{U}_{\tilde{s}}^\dagger
=
\hat{U}_{t}^\dagger 
e^{i \frac{g_0}{z_{\text{zpf}}}\hat{z}_t (t-\tilde{s})}
\label{uapprox}
\end{equation}
Using then Eq.\eqref{uapprox} in Eq.~\eqref{alpha1} we obtain:
\begin{equation}
\hat{a}_t= E  \int_{t-\frac{2}{\kappa}}^t e^{-\frac{\kappa}{2}(t-\tilde{s})} e^{i\frac{g_0}{z_\text{zpf}}\hat{z}_t (t-\tilde{s})}  d\tilde{s},
\label{alpha2}
\end{equation}
where we have also assumed that $\hat{a}_0\approx 0$, i.e. the cavity mode is initially not populated. We finally approximate the integral in Eq.~\eqref{alpha2} by the mean value and obtain:
\begin{equation}
\hat{a}_t= 2e^{-\frac{1}{2}}\frac{E}{\kappa} e^{i\frac{g_0}{\kappa z_{\text{zpf}}}\hat{z}_t} 
\label{alpha3}
\end{equation}

Comparing Eq.~\eqref{alpha3} with Eq.~\eqref{Es} we find the following relation:
\begin{equation}
\frac{g_0}{\kappa}= 4\pi\frac{z_{\text{zpf}}}{\lambda},
\label{g0kappa}
\end{equation}
where, on the left hand-side, we have cavity system quantities, and, on the right hand-side, we have free space system quantities. Specifically, inserting the values of the free space system presented in the main text in Eq.~\eqref{g0kappa} we find $g_0/\kappa \approx 10^{-4}$. This ratio is important for the discussion of non-linear quantum opto-mechanics~\cite{aspelmeyer2014cavity}. We leave a more refined, fully quantum mechanical description of free space systems, and the comparison with cavity systems, for future research.

{\bf Detection schemes--}
Comparing Eq.~\eqref{alpha3} with Eq.~\eqref{Es} also shows that we can monitor the mechanical motion in real-time~\cite{tufarelli2012input}. In particular, the CH, CBH detection schemes, discussed in the main text, can be applied also a cavity opto-mechanical experimental setup. Specifically, the analysis from Eq.~\eqref{sl} to Eq.~\eqref{Ns_approx3} remains valid.

\end{document}